\documentstyle[12pt]{article}
\newcommand {\bra}[1]{\langle#1\vert}
\newcommand {\ket}[1]{\vert#1\rangle}
\begin{document}
\centerline{{\huge Analytic Solutions to the Problem of}}
\medskip
\centerline{{\huge  Coulomb and Confining Potentials}}
\medskip
\centerline{M. Dineykhan and R.G. Nazmitdinov}
\medskip
\centerline{Bogoliubov Laboratory of Theoretical Physics}
\centerline{Joint Institute for Nuclear Research, 141980 Dubna, Russia}
\baselineskip 20pt minus.1pt
\begin{abstract}
The oscillator representation method is presented and used
to calculate the energy spectra of a superposition of
Coulomb and power-law potentials and for Coulomb and Yukawa potentials.
The method provides an efficient way to obtain analytic results 
for arbitrary values of the parameters specifying the above-type 
potentials. The calculated energies of the ground and excited states of
the quantum systems in question  are found to comply well
with the exact results.
\end{abstract}
\vspace{0.2in}

\vspace{0.2in}
\section{Introduction}

The combination of Coulomb potential with different type of
confining potentials is a typical problem in
quantum mechanics for many applications.
The two-dimensional electron gas
with Coulomb interaction confined in a parabolic potential
is a model of a quantum dot (see for example \cite{Tau,DN} and
references therein), i.e. an artificial atom
the properties of which can be controlled by a man (for review on
quantum dots see \cite{Kast,Jon}). The combination of
Coulomb and Yukawa potentials, called
the exponentially screened Coulomb potential (ESCP),
describe the effective two-particle interaction for
charged particles in an ionized gas or in a
metal (see \cite{Ad} and references therein).
The model of quark confinement
of charmed quarks is successfully simulated by a simple
superposition of the linear and Coulomb term \cite{Cor}.

Whereas many approximate
analytical and numerical methods have been developed
for the last years for the analysis of such a superposition
of potentials \cite{AFC}, we would like
to suggest a method which is called the oscillator
representation method (ORM) \cite{Din}.
The ORM, based on the ideas and methods of quantum field theory,
has been proposed to calculate the binding energy of different systems with
fairly arbitrary potentials described by the Schr\"{o}dinger equation (SE).
A spectrum of the one-dimensional power--law potential
or of the quartic oscillator, which are test benches
for approximate methods (see for example \cite{Vor,BL}),
can be determined analytically in the ORM with accuracy
very closed to the best, available nowadays, computational methods
(see for the comparision \cite{Din}).
One of our goals is to present the mathematical formalism of the ORM
in detail making it available
for further applications, in particular, for calculations of ground and
excited states of the systems with Coulomb interaction.

In situation, when different terms of
the effective potential of a system can compete, and
there is no basis for the perturbative approach to be applied,
a solution of the SE becomes a nontrivial problem.
However, the lowest levels of such a system, especially ones
closest to the ground state, can be described
with a good approximation by the oscillator states
defined in the expanded hyperspace  which includes the original
one. For the first time, Fock demonstrated this
result for a nonrelativistic hydrogen atom
by making a stereographic projection of the momentum space
onto a four--dimensional unit sphere \cite{Fock}.
For an arbitrary smooth potential, admitting the existence
of system bound states, its eigenfunctions have exponential
asymptotic behavior at large distances \cite{LL,Mes}.
Therefore, the variables in the original SE must
be changed so that the modified equation should have solutions
with  the oscillator behavior at large distances.
Schr\"{o}dinger pointed out at the existence of
such a transformation which turned the three-dimensional Coulomb
system into the oscillator one in the four-dimensional space \cite{Sc}.
The explicit form of this transformation has been found in \cite{Kus}
and used to solve the classical Kepler problem.
Since this transformation is not a canonical one, after the
transformation we have a new system
with another set of quantum numbers and wave functions
which contains, however, a subset of the original wave functions.
The transformation of variables, leading to
the Gaussian asymptotic behavior
in the expanded space, is one of the basic elements of the ORM.

As the next step, it is necessary to represent the canonical variables
(coordinate and momentum) of the Hamiltonian through
the creation and annihilation operators $a^+$ and $a$.
The pure oscillator part with some, yet unknown,
frequency $\omega$ is extracted from the Hamiltonian, i.e.
$H\rightarrow H_0 + H_{I} = \omega a^+ a + higher~~ order~~ terms$.
The remaining part, i.e.
the interaction Hamiltonian $H_I$, is represented in terms of normal products
over $a^+$ and $a$. In addition, it is required that the interaction
Hamiltonian does not contain terms quadratic in the canonical variables.
This condition is equivalent to the equation
\begin{equation}
\label{cond}
\frac{d\varepsilon_0}{d\omega} = 0
\end{equation}
which determines $\omega$, the oscillator frequency, in the ORM
and is called the {\it oscillator representation condition} (ORC) \cite{Din}.
Similar ideas are used in the Hartee-Fock-Bogoliubov theory to
describe different correlations between nucleons moving in an average
nuclear potential (see for review \cite{Sol}), in order to include
particle correlations into a new quasiparticle vacuum.
In fact, the ORC coincides
with the variational equation arising after averaging of the total
Hamiltonian over the Gaussian probe functions in the expanded
configuration space. In addition, the dimension of
the hyperspace is the variational parameter in the ORM.
Using the ORM, we will calculate the energy spectrum and wave functions
for Coulomb and power--law potentials and for
the exponentially screened Coulomb potential.

The paper is organized as follows: In Sect. 2, basic formulae of the ORM
for spherical potentials are presented. In Sect. 3 the method is applied
to calculate the energy spectrum for Coulomb and power-law potentials,
for Coulomb and Yukawa potentials. In Sect. 4 we discuss the main
results. In the appendices, useful
relations used in the ORM are presented.

\section{ORM in the Space $R^D$}

Let us consider the Schr\"{o}dinger equation in
3--dimensions
\begin{equation}
\label{gen}
\int d^3 {\bf r}\Psi ({\bf r})\left[-\frac{1}{2}\Delta + V(r)-E\right]
\Psi ({\bf r}) = 0
\end{equation}
where $V(r)$ is
a potential of Coulomb and/or Yukawa type, decreasing at
large distances ($V(r)\rightarrow 0~~~ {\rm as}~~~ r\rightarrow\infty$).
If $\Psi ({\bf r})=\psi_{nl}(r) Y_{nl}(\theta , \phi )$, this equation
for the wave function of the $l$-th orbital excitation is \cite{LL,Mes}
\begin{eqnarray}
\label{d31}
\int_{0}^{\infty}dr(r\psi_{nl}(r) )\left[-\frac{1}{2}{\Big(}\frac{d}{dr}{\Big)}^2
+\frac{l(l+1)}{2r^2}
+V(r)-E\right](r\psi_{nl}(r))=0
\end{eqnarray}
where the function $\psi_{nl}(r)$ depends on a radial variable only.
General discussion on the transition to arbitrary dimensions in the
radial SE is given in \cite{Jon1}. In this case, calculation of
the wave function $\psi_{0l}(r)$ would be equivalent to that
of the ground state wave function of the modified Hamiltonian in
another dimension. The radial excitation wave functions
$\psi_{nl}(r)=\ket{n}$  $(n=n_r + l +1)$
are equivalent to the highest oscillator states.

At large distances, for Coulomb and Yukawa type
potentials the asymptotic behaviour of the wave function $\psi(r)$
as $r\rightarrow\infty$ is well known, i.e. $\psi(r)\sim\exp(-a(r))~~$.
The new variable $r=r(q)$ should be introduced so that
$a(r(q))~\sim~q^2~~~~~{\rm as}~~~~~q\to\infty$
and $\psi(r(q))~\sim~\exp(-q^2).$
Using the substitutions
\begin{eqnarray}
\label{subs}
r=q^{2\rho} ~~~~~~{\rm and}~~~~~r\psi_{0l}(r)=q^a\phi(q),
\end{eqnarray}
after simple calculations we obtain
\begin{eqnarray}
\label{d33}
\int_{0}^{\infty}dq q^{D-1}\phi(q){\Big[}-{1\over 2}{\Big(}\left( {d\over dq}
\right)^2+{{D-1}\over q}{d\over dq}{\Big)}
+W_{I}(q^2;E){\Big]}\phi(q)=0~~,
\end{eqnarray}
where
\begin{equation}
\label{pot0}
W_{I}(q^2;E) = -{K(l,\rho,D)\over 2q^2}+4{\rho}^2(q^2)^{(2\rho-1)}(V(q^2)-E),
\end{equation}
\begin{equation}
\label{dim}
D = 2a-2\rho+2
\end{equation}
\begin{equation}
K(l,\rho,D)={1\over 4}{\Big(}(D-2)^2-4\rho^2(2l+1)^2{\Big)}
\end{equation}
We can identify Eq.(\ref{d33}) with the SE in
the expanded space ${\rm R}^D$ for the wave function $\phi(q)$
depending on the radial variable $q$ only
\begin{eqnarray}
\label{d36}
\int d^D{\bf q}\phi({\bf q}){\Bigg[}-{1\over2}\Delta_D+W_I(q^2;E)-
\varepsilon(E){\Bigg]}\phi({\bf q})=0,
\end{eqnarray}
where the function
\begin{equation}
\varepsilon(E)=\varepsilon(l,\rho,D;E)
\end{equation}
should be considered as an eigenvalue of the SE in D--dimensions
\begin{equation}
\label{d37}
{\Big[}-{1\over2}\Delta_D+W_I(q^2,E){\Big]}\phi(q)=\varepsilon(E)\phi(q)
\end{equation}
We note that the potential $W_I(q^2;E)$ contains
the attractive term which can compensate the repulsive part
of the $V(q^2)$.

The energy spectrum $E_{n\ell}$ of the original system
is contained in the radial excitation spectrum
$\varepsilon^{[n_r]}$ of the Hamiltonian of Eq.({\ref{d37})
\begin{eqnarray}
\label{d38}
&&H(E)\phi^{[n_r]}(q)=\varepsilon^{[n_r]}(E)\phi^{[n_r]}(q)~,~~(n_r=0,1,2,...).
\end{eqnarray}
and it is determined by the equation
\begin{eqnarray}
\label{d39}
&&\varepsilon^{[n_r]}(E)=\varepsilon^{[n_r]}(\ell,\rho,D;E)=0.
\end{eqnarray}

The additional parameters $\rho$ and $D$ can be found,
for example, by the minimization
of the energy in the zeroth approximation
$$\varepsilon_0(E)=\min_{\{\rho,D\}}\varepsilon(l,\rho,D;E).$$
On the other hand, the parameter $\rho$ is connected
with the condition of the Gaussian asymptotic behavior for
the wave function (see the discussion above).
The parameter $D$ is connected with the behaviour
of the wave function at short distances.
If the potential $V(r)$ has no a repulsive character
as $r\to 0$ in the ORM we choose $K(l,\rho,D)=0$
and
\begin{eqnarray}
\label{t1}
D = 2+2\rho(2l+1).
\end{eqnarray}
In the opposite case, the parameter $D$ is chosen
to suppress a repulsion produced by the potential
$V(r)$ at $r\to 0$. In fact, the parameter D is used
to improve the zeroth approximation at the minimization
of the ground state energy.

One can see that the radial quantum number $n_r$ does not enter into
Eq.(\ref{d37}) in an explicit form.
The orbital quantum number $\ell$ is absorbed by the parameter $D$.
From the point of view of the expanded space ${\rm R}^D$ the functions
$$\phi_n(q)=q^{2\rho-a}\psi_{nl}(q^{2\rho})~~~~~~{\rm or}~~~~~~
\psi_{n\ell}(r)=r^{{D-2\rho-2\over 4\rho}}\phi_n(r^{1/(2\rho)})$$
for any $n$ and for fixed $\ell$ are eigenfunctions
of the basic series of radial excitations in the space
${\rm R}^D$ with radial quantum number
$n$ and zeroth orbital momentum. Consequently, the solution of the
SE in 3 dimensions for $\ell$ orbital excitation is
equivalent to the solution of the SE in the space ${\rm R}^D$
for states with the zeroth angular momentum.
As a result, the original SE is represented in the form of
Eq.(\ref{d37}) in which the wave function of the ground state
$\phi(q)$ has
\begin{itemize}
\item a Gaussian asymptotic behaviour at large distances
$\phi(q)\sim \exp(-q^2)$,
\item a maximum at the point $q=0$.
\end{itemize}
In addition, the integration over the expanded space
$R^D$ and the Laplacian are defined by:

   $$i)~~~~\int d^D{ \bf q}=\frac{2\pi^{D/2}}{\Gamma(D/2)}
   \int\limits_{0}^{\infty}dq\cdot q^{D-1}$$
   $$ii)~~~~\Delta_q=\left(\frac{d^2}{dq^2}+\frac{D-1}{q}\cdot\frac{d}{dq}
   \right).$$
\vspace{.5cm}

The next step is to introduce the oscillator representation which
can be done in the following way.
The Hamiltonian $H$ can be rewritten  in the form
\begin{eqnarray}
\label{2.6}
H=\frac{1}{2}(p^2+\omega^2q^2)
+{\Big(}W(q)-\frac{1}{2}\omega^2q^2 {\Big)}~,
\end{eqnarray}
where $\omega$ is an unknown oscillator frequency.
Let us substitute the creation and annihilation operators
(Appendix A) into Eq.(\ref{2.6}) and go
to the normal product of the operators $a^{+}_j$ and $a_j$.
Hereafter we use the notation $d\equiv D$. One can
obtain
\begin{eqnarray}
\label{2.8}
&&\frac{1}{2}(p^2+\omega^2q^2)=\omega\sum_j a_j^{+}a_j+\frac{d}{2}\omega
=\omega (a^{+}a)+\frac{d}{2}\omega,\\
&&W(q)-\frac{\omega^2}{2}q^2=\int{\Big(}{dk\over2\pi}{\Big)}^d
{\widetilde{W}}_d(k^2)\exp{\Big(}-\frac{k^2}{4\omega}{\Big)}:e^{i(kq)}:
-\frac{\omega^2}{2}{\Big (}:q^2:+\frac{d}{2\omega}{\Big )}~, \nonumber
\end{eqnarray}
where $:*:$ is the symbol of the normal ordering and $(qk)=\sum_j k_jq_j$,
\begin{equation}
\label{fur}
{\widetilde{W}}_d(k^2)=\int(d\rho)^dW(\rho)e^{i(k\rho)}~.\nonumber
\end{equation}
We require that the interaction part of the Hamiltonian should not contain
the term with $:q^2:$ because this term is postulated to be included into the
oscillator part completely. This requirement gives the equation for the
frequency $\omega$:
\begin{eqnarray}
\label{2.9}
\omega^2-\int{\Big(}{dk\over2\pi}{\Big)}^d{\widetilde{W}}_d(k^2)
\exp(-\frac{k^2}{4\omega})\frac{k^2}{d}=0~.
\end{eqnarray}
Using these equations, we can rewrite the Hamiltonian, Eq.(\ref{2.6}),
in the form:
\begin{equation}
\label{2.10}
H = H_0+H_I+\varepsilon_0~,
\end{equation}
\begin{equation}
H_0=\omega(a^{+}a)~,
\end{equation}
\begin{eqnarray}
H_I &=& \int{\Big(}{dk\over2\pi}{\Big)}^d
{\widetilde{W}}_d(k^2)\exp(-\frac{k^2}{4\omega})
:e^{i(kq)}-1+\frac{k^2q^2}{2d}:\nonumber\\
~&=&\int{\Big(}{d\rho\over\sqrt{\pi}}{\Big)}^de^{-\rho^2}
W(\frac{\rho}{\sqrt{\omega}})
:\exp(-q^2+2(\rho q))-1+q^2(1-\frac{2\rho^2}{d}):~.
\end{eqnarray}
\begin{equation}
\label{eps0}
\varepsilon_0 = \frac{d\omega}{4}+\int{\Big(}{dk\over2\pi}{\Big)}^d
{\widetilde{W}}_d(k^2)\exp(-\frac{k^2}{4\omega}).
\end{equation}
Details of calculations of different integrals appearing in the above
expressions can be found in Appendices.
In the ORM the solution of the SE Eq.(\ref{d37}) has the form
$$\phi_n(q)=\exp(-{\omega\over2}q^2)\sum_m c_{nm}P^{(D)}_m(q^2\omega),$$
where ${\{}P^{(D)}_m(t){\}}$  is the class of orthogonal polynomials
which are orthogonal within the interval $0<t<\infty$  with the weight
function
$$\rho_D(t)=t^{{D\over2}-1}\exp(-t),$$
i.e.,
\begin{eqnarray}
\int\limits_{0}^{\infty}dt~t^{{D\over2}-1}{\rm e}^{-t}P^{(D)}_n(t)
P^{(D)}_m(t)=\delta_{nm}~.
\nonumber
\end{eqnarray}
These orthogonal polynomials can be constructed
by using the formalism of creation and annihilation operators
$a_j$ and $a^+_j$  in the space  ${\rm R}^D$ (see Appendices A, B).
Using the definition for the ground wave function in
the $R^D$ space
\begin{eqnarray}
\label{gr}
\ket{0}=\prod_{j=1}^{D}\frac{\omega^{1/4}}{\pi^{1/4}}{\rm e}^
{-\frac{\omega}{2}q^2_j}=
{\Big(}\frac{\omega}{\pi}{\Big)}^{D/4}{\rm e}^{-\frac{\omega}{2}q^2}
\end{eqnarray}
we can write for radial excitations $n = 1,2,...$
\begin{eqnarray}
\label{rad}
\Phi_n&\sim&(a^{+}a^{+})^n\ket{0}
\sim P^{(D)}_n(\omega q^2){\rm e}^{-{\omega\over2}q^2}\\
&\sim& P^{(D)}_n(\omega r^{1/\rho})\exp(-{\omega\over2}r^{1/\rho}),
\nonumber
\end{eqnarray}
where  $P^{(D)}_n(t)$ is a polynomial of the $n$ th order.
The parameter $D$ in this representation can be considered
to be any positive number. These polynomials satisfy the orthogonal
condition
\begin{eqnarray}
&&{\Big(}\Phi_n,\Phi_m {\Big)}\sim\bra{0}(aa)^n(a^{+}a^{+})^m\ket{0}
\nonumber\\
&&\sim\int\limits_{0}^{\infty}dq~q^{D-1}
\exp(-q^2)P^{(D)}_n(q^2)P^{(D)}_m(q^2) \nonumber\\
&&\sim\int\limits_{0}^{\infty}dt~t^{D/2-1}
\exp(-t)P^{(D)}_n(t)P^{(D)}_m(t) \sim\delta_{nm}~.\nonumber
\end{eqnarray}

Thus, the ORC, Eq.(\ref{cond}), is written as
\begin{eqnarray}
\label{2.11}
\frac{\partial}{\partial{\omega}}\varepsilon_0(E;\omega,d)=0~.
\end{eqnarray}
This equation determines the parameter
$\omega=\omega(E,d)~$
as a function of the energy $E$, $d$ and other parameters defining the
potential $V(r)$ in Eq.(\ref{d31}).

 According to Eq.(\ref{d39}), the ground state energy $E$ of the original
problem in the $N$-th perturbation order of the ORM is
defined by Eq.(\ref{2.11}) and by the equation
\begin{eqnarray}
\label{2.12}
&&\varepsilon_{(N)}(E,d)=\varepsilon_0(E,d)+\varepsilon_2(E,d)+. . .
+\varepsilon_N(E,d)=0~.
\end{eqnarray}
Eq.(\ref{2.12}) determines the energy $E_{(N)}(d)$ in the $N$-th perturbation
order as a function of $d$ or $\rho$ and other parameters defining the
potential. The function $\varepsilon(E;\omega,d)$ in Eq.(\ref{2.10})
depends on two
parameters $d$ and $\omega$. Let us denote the parameter $d$ and
other additional parameters by ${\{}\alpha_j{\}}$.
The ground state energy in the expanded space up to second order
is
\begin{eqnarray}
&&\varepsilon_{(2)}(E;\omega,\alpha_j)=\varepsilon_0(E;\omega,\alpha_j)+
\varepsilon_2(E;\omega,\alpha_j)~. \nonumber
\end{eqnarray}
According to Eqs.(\ref{eps0}), (\ref{fur}) and (\ref{pot0}), the function
$\varepsilon_0(E;\omega,\alpha_j)$ can be written
in the form
\begin{equation}
\label{zer}
\varepsilon_0(E;\omega,\alpha_j)=A(\omega,\alpha_j)-E\cdot
B(\omega,\alpha_j),
\end{equation}
where $A(\omega,\alpha_j)$ and $B(\omega,\alpha_j)$
are known functions.

In the zeroth order Eqs.(\ref{2.12}) and (\ref{zer}) lead to equations
\begin{equation}
\label{z1}
{\partial\over\partial\alpha_k}E_0(\omega,\alpha_j)
={\partial\over\partial\alpha_k}
\left({A(\omega,\alpha_j)\over B(\omega,\alpha_j)}\right)=0
~~~~~~~{\rm for~~ all}~~k,
\end{equation}
which determine the parameters ${\{}\alpha_j{\}}$ as functions
of $E_0$
\begin{equation}
\label{z2}
\alpha_j=\alpha_j(E).
\end{equation}
As a result, in the zeroth approxiamtion we have
\begin{equation}
\label{z3}
E_0=\min_{{\{}\omega,\alpha_j{\}}}{A(\omega,\alpha_j)
\over B(\omega,\alpha_j)}={A(\omega_0,\alpha_j^0)\over
B(\omega_0,\alpha_j^0)}
\end{equation}
where the parameters $\omega_0$ and $\alpha_j^0$ determine the minimum.

In the second approximation we solve the following equation
\begin{equation}
\label{z4}
\varepsilon_{(2)}(E;\omega,\alpha_j)=\varepsilon_0(E;\omega,\alpha_j)+
\varepsilon_2(E;\omega,\alpha_j)=0~.
\end{equation}
We assume that the second correction is small one and
the energy is $E_{(2)}=E_0+E_2$. Consequently, we have
\begin{equation}
\varepsilon_0(E_{(2)};\omega_0,\alpha_j^0)
=A(\omega_0,\alpha_j^0)-E_{(2)}B(\omega_0,\alpha_j^0),
\nonumber
\end{equation}
where $\omega_0=\omega(E_0)$.
Therefore, the second correction is
\begin{eqnarray}
&&E_2={\varepsilon_2(E_0;\omega_0,\alpha_j^0)\over B(\omega_0,\alpha_j^0)}
+O(E^2_2). \nonumber
\end{eqnarray}
Finally, we have
\begin{eqnarray}
\label{1,2.23}
E_{(2)}&=&E_0+E_2= \\
&=&\min_{{\{}\omega,\alpha_j{\}}}
{A(\alpha_j)\over B(\alpha_j)}+
{\varepsilon_2(E_0;\omega_0,\alpha_j(E_0))\over B(\omega_0,\alpha_j(E_0))}
+E_0O\left(\left|{E_2\over E_0}\right|^2\right) \nonumber\\
&=&{A(\omega_0,\alpha_j^0)+\varepsilon_2(E_0;\omega_0,\alpha_j^0)\over
B(\omega_0,\alpha_j^0)}+E_0O\left(\left|{E_2\over E_0}\right|^2\right)~~.
\nonumber
\end{eqnarray}
In our calculations we use essentially these equations.
The accuracy of the oscillator representation can be evaluated as
\begin{eqnarray}
\delta\sim\left|{\varepsilon_2\over\varepsilon_0}\right|~.
\nonumber
\end{eqnarray}

For the radial excitations $\ket{n_r}$
$({n_r}=1,2,...)$ the matrix element
\begin{eqnarray}
\bra{n_r}H_I\ket{n_r}=A^{[n_r]}(\omega,\alpha_j)-
EB^{[n_r]}(\omega,\alpha_j)\neq0~. \nonumber
\end{eqnarray}
The energy $\varepsilon^{[n_r]}$ in the lowest approximation looks like
\begin{eqnarray}
\label{2.14}
\varepsilon_1^{[n_r]}(E)&=&\bra{n_r}H\ket{n_r}=\varepsilon_0(E)+2{n_r}\omega+
\bra{n_r}H_I\ket{n_r}\nonumber\\
&=&A_1^{(n_r)}(\omega,\alpha_j)-EB_1^{(n_r)}(\omega,\alpha_j)~,
\end{eqnarray}
where
\begin{eqnarray}
&&A_1^{({n_r})}(\omega,\alpha_j)=
A(\omega,\alpha_j)+2{n_r}\omega+A^{[n_r]}(\omega,\alpha_j)~,\nonumber\\
&&B_1^{(n_r)}(\omega,\alpha_j)=B(\omega,\alpha_j)+B^{[n_r]}(\omega,\alpha_j)~.
\nonumber
\end{eqnarray}
Two equations
\begin{eqnarray}
\label{2.15}
&&\frac{\partial}{\partial{\omega}}A(\omega,\alpha_j)-
E\frac{\partial}{\partial{\omega}} B(\omega,\alpha_j)=0~,\\
&&A_1^{(n_r)}(\omega,\alpha_j)-EB_1^{(n_r)}(\omega,\alpha_j)=0~, \nonumber
\end{eqnarray}
determine the functions $\omega(\alpha_j)$ and $E(\alpha_j)$.
The energy of the $n_r$-th excited state in the first approximation
of the oscillator representation is determined as
\begin{eqnarray}
&&E^{[n_r]}=\min_{\{\alpha_j\}}{A_1^{[n_r]}(\omega(\alpha_j),\alpha_j)\over
B_1^{[n_r]}(\omega(\alpha_j),\alpha_j)}. \nonumber
\end{eqnarray}

Further steps should be done according to the rules formulated above.

\section{Coulomb type potentials}

\subsection{Coulomb and power--law confining potentials}

In the general case, the potential under consideration can be written
as follows
\begin{eqnarray}
\label{k1}
V(r)=-\frac{1}{r}+g\cdot r^\nu~.
\end{eqnarray}
At $\nu=1$ this potential corresponds to the well-known "Cornell"
potential \cite{Cor}, used
to study the energy spectrum of a many-quark system.
In addition, at $\nu=1$ and 2 the potential can be used
to study the Stark problem  and the Zeeman effect
in a spherical system, respectively.
Due to various applications of this kind of potential, there
are many attempts to solve
the eigenvalue problem with a good quantitative accuracy
(see, for example \cite{AFC,FF,GS} and references
therein). In fact, the main approach is based on ideas of
the perturbation theory. As a result, the problem of summing
the perturbation series is not solved completely.

We apply the ORM to calculate the energy spectrum
for an arbitrary value of the parameter $\nu$.
Using the transformation Eq.(\ref{subs}), after some simple
calculations the SE with the potential Eq.(\ref{k1}) can be written
in the following form
\begin{eqnarray}
\label{k2}
&&\left[-\frac{1}{2}\left(\frac{{\partial}^2}{{\partial}q^2}
+\frac{d-1}{q}\frac{{\partial}}{{\partial}q}\right)-4\rho^2Eq^{2(2\rho-1)}
-4\rho^2q^{2(\rho-1)}\right.\\
&&\left.+4\rho^2gq^{2(\nu\rho+2\rho-1)}\right]\Phi(q)=
\varepsilon(E)\Phi(q)~.
\nonumber
\end{eqnarray}
\begin{equation}
d=2+ 2\rho + 4\rho l
\end{equation}
According to the ORM, the Hamiltonian has the form of Eq.(\ref{2.10})
where
\begin{eqnarray}
\label{k3}
&&H_0=\omega\left(a^+a\right)~,\\
&&H_I=\int\limits_{0}^{\infty}dx\int\left(\frac{d\eta}
{\sqrt{\pi}}\right)^de^{-\eta^2(1+x)}:e_2^{-2i\sqrt{x\omega}(q\eta)}:
\nonumber\\
&&\left[-\frac{4\rho^2E}{\omega^{2\rho-1}}
\frac{x^{-2\rho}}{\Gamma(1-2\rho)}
-\frac{4\rho^2}{\omega^{\rho-1}}
\frac{x^{-\rho}}{\Gamma(1-\rho)}
+\frac{4g\rho^2}{\omega^{\nu\rho+2\rho-1}}
\frac{x^{-2\rho-\nu\rho}}{\Gamma(1-2\rho-\nu\rho)}
\right]
\nonumber
\end{eqnarray}
and the function $\varepsilon_0$ is
\begin{eqnarray}
\varepsilon_0(E)=\min_{\{\omega,\rho\}}\varepsilon_0(E;\omega,\rho)~,
\nonumber
\end{eqnarray}
Here
\begin{eqnarray}
\label{k4}
&&\varepsilon_0(E;\omega,\rho)={d\omega\over 4}
-\frac{4E\rho^2}{\omega^{2\rho-1}}
\frac{\Gamma(d/2+2\rho-1)}{\Gamma(d/2)}\\
&&-\frac{4\rho^2}{\omega^{\rho-1}}
\frac{\Gamma(d/2+\rho-1)}{\Gamma(d/2)}
+\frac{4g\rho^2}{\omega^{2\rho+\nu\rho-1}}
\frac{\Gamma(d/2+2\rho+\nu\rho-1)}{\Gamma(d/2)}
\nonumber
\end{eqnarray}
The energy of the system and the new oscillator frequency are
determined from Eq.(\ref{d39})
and  from the ORC, Eq.(\ref{2.11}), respectively .
After some simplifications, we obtain the following equation
from the system of two equations:
\begin{eqnarray}
\label{k5}
E&=&\min_{[\rho]}
\left\{
\frac{Z^2}{8\rho^2}
\frac{\Gamma(d/2+1)}{\Gamma(d/2+2\rho-1)}
-\frac{Z\Gamma(d/2+\rho-1)}{\Gamma(d/2+2\rho-1)}
\right.\nonumber\\
&+&\left. \frac{g}{Z^\nu}
\frac{\Gamma(d/2+2\rho+\nu\rho-1)}{\Gamma(d/2+2\rho-1)}
\right\}~,
\end{eqnarray}
The parameter $Z$ is determined from Eq.(\ref{2.11})
\begin{eqnarray}
\label{k6}
Z^{2+\nu} &-&Z^{1+\nu}4\rho^2
\frac{\Gamma(d/2+\rho-1)}{\Gamma(d/2+2\rho-1)} \\
&-& g\nu4\rho^2
\frac{\Gamma(d/2+2\rho+\nu\rho-1)}{\Gamma(d/2+2\rho-1)}=0
\nonumber
\end{eqnarray}
and the oscillator frequency is
$$\omega=Z^{1/\rho}~.$$

In the strong coupling limit, i.e. as $g\to\infty$,
we obtain from Eq.(\ref{k5}) the following result
\begin{eqnarray}
\label{k7}
E=g^{\frac{2}{2+\nu}}\cdot C~,
\end{eqnarray}
where
\begin{eqnarray}
\label{k8}
&&C=\min_{[\rho]}
 \left\{\left(\frac{1}{2}+\frac{1}{\nu}\right)
\right.\\
&&\left.
\frac{\Gamma(d/2+1)}{4\rho^2\Gamma(d/2+2\rho-1)}
\left[4\nu\rho^2
\frac{\Gamma(d/2+2\rho+\nu\rho-1)}{\Gamma(d/2+1)}
\right]^{\frac{2}{2+\nu}}\right\}~.
\nonumber
\end{eqnarray}

In Table 1 we present, as an illustration, the results
at the zeroth approximation of the ORM for $\nu=1$.

\subsection{The exponentially screened Coulomb potential}

The ESCP describes various physical phenomena.
It was used to evaluate the cohesive energy of alkali
metals \cite{Hel}, to calculate of energy bands of sodium
by empirical pseudopotential method \cite{Cal} and to
describe of the exciton and donor spectra in ionic crystals
and polar semiconductors \cite{Pol,ad1}.

The Schr\"{o}dinger equation for the ESCP has the following form:
\begin{eqnarray}
\label{s1}
\left[\frac{1}{2\mu}p^2-\frac{A}{r}+B\cdot\frac{e^{-cr}}{r}\right]
\Psi(r)=E\Psi(r)~,
\end{eqnarray}
where $\mu$ is relative mass, $A$, $B$ and $c$ are some constants.
We use the following units: the length unit is
$\hbar^2/(A\mu)$ and the energy unit is
$\mu A^2/(2\hbar^2)$.  The SE in the dimensionless units can be written
\begin{eqnarray}
\label{s2}
\left[-\frac{1}{2}\Delta-\frac{1}{r}+
\frac{B}{2} \frac{e^{-cr}}{r}-\frac{E}{2}\right]
\Psi(r)=0
\end{eqnarray}

Using the transformation Eq.(\ref{subs}), we obtain the modified SE
\begin{eqnarray}
\label{s3}
&&\left[-\frac{1}{2}\left(\frac{{\partial}^2}{{\partial}q^2}
+\frac{d-1}{q}\frac{{\partial}}{{\partial}q}\right)-2\rho^2Eq^{2(2\rho-1)}
-4\rho^2q^{2(\rho-1)}\right.\\
&&\left.+2B\rho^2q^{2(2\rho-1)}
\exp\{-cq^{2\rho}\}\right]\Phi(q)=
\varepsilon(E)\Phi(q)~;\nonumber \\
&&\varepsilon(E)=0~,~~~~~~~~
d=2+2\rho+4\rho\ell~.
\nonumber
\end{eqnarray}
Applying the ORM to the Hamiltonian Eq.(\ref{s3}),
we obtain
$$H=H_0+H_I+\varepsilon_0,$$
where
\begin{eqnarray}
\label{s4}
H_0&=&\omega\left(a^+a\right)~,\\
H_I&=&\int\limits_{0}^{\infty}dx\int\left(\frac{d\eta}
{\sqrt{\pi}}\right)^de^{-\eta^2(1+x)}:e_2^{-2i\sqrt{x\omega}(q\eta)}:
\nonumber\\
&&\left[-\frac{-2\rho^2E}{\omega^{2\rho-1}}
\frac{x^{-2\rho}}{\Gamma(1-2\rho)}
-\frac{4\rho^2}{\omega^{\rho-1}}
\frac{x^{-\rho}}{\Gamma(1-\rho)}
\right] \nonumber\\
&&+2B\rho^2\sum_{k=0}^{\infty}\frac{(-c)^k}{k!\omega^{k\rho+\rho-1}}
\int\limits_{0}^{\infty}dx\frac{x^{-\rho(1+k)}}{\Gamma(1-\rho(1+k))}
\nonumber\\
&&\int\left(\frac{d\eta}
{\sqrt{\pi}}\right)^de^{-\eta^2(1+x)}:e_2^{-2i\sqrt{x\omega}(q\eta)}:
\nonumber\\
\label{s5}
\varepsilon_0(E)&=&{d\omega\over 4}
-\frac{2E\rho^2}{\omega^{2\rho-1}}
\frac{\Gamma(d/2+2\rho-1)}{\Gamma(d/2)}
-\frac{4\rho^2}{\omega^{\rho-1}}
\frac{\Gamma(d/2+\rho-1)}{\Gamma(d/2)}
\nonumber\\
&+&\frac{2B\rho^2}{\Gamma(d/2)}
\int\limits_{0}^{\infty}du
u^{\frac{d}{2}-1}e^{-u}\left(\frac{u}{\omega}\right)^{\rho-1}
\exp\{-c\left(\frac{u}{\omega}\right)^{\rho}\}~.
\end{eqnarray}
The energy $E$ and the new oscillator frequency of the system are
determined from Eq.(\ref{d39}) and Eq.(\ref{2.11}), respectively.
After some algebraic calculations, the solution of the system of these
two equations has the form
\begin{eqnarray}
\label{s6}
E&=&\min_{[\rho]}
\left\{
\frac{Z^2}{4\rho^2}
\frac{\Gamma(2+\rho+2\rho\ell)}{\Gamma(3\rho+2\rho\ell)}
-2Z\frac{\Gamma(2\rho+2\rho\ell)}{\Gamma(3\rho+2\rho\ell)}
\right.\\
&+&\left.\frac{ZB}{\Gamma(3\rho+2\rho\ell)}
\int\limits_{0}^{\infty}du
u^{2\rho+2\rho\ell-1}
\exp\{-u-\frac{c}{Z}u^\rho\}
\right\}~,
\nonumber
\end{eqnarray}
Here the parameter $Z$ is determined from Eq.(\ref{2.11})
\begin{eqnarray}
\label{s7}
&&Z^2-4Z\rho^2
\frac{\Gamma(2\rho+2\rho\ell)}{\Gamma(2+\rho+2\rho\ell)}\\
&+&\frac{2B\rho^2}{\Gamma(2+\rho+2\rho\ell)}
\int\limits_{0}^{\infty}du
u^{2\rho+2\rho\ell-1}\left(Z+cu^\rho\right)
\exp\{-u-\frac{c}{Z}u^\rho\}
=0
\nonumber
\end{eqnarray}
and the oscillator frequency is
$$\omega = Z^{1/\rho}~.$$
In Tables 2 and 3 we present the result of calculations of the ESCP energy
for different sets of parameters. The comparision with the computational
results \cite{Ad} is very good and it gives a strong support to
the ideas of the ORM.

\section{Summary}

We presented the main ideas  and the mathematical formalism
of the ORM which is very efficient in getting the analytical
solution to bound states of a stationary quantum--mechanical problem.

Using the ORM, the discrete eigenvalues for
the superposition of Coulomb and power-law potentials
have been calculated. The results are obtained
for arbitrary values of the strength and the degree
of the power-law potential. It is a well-known fact that
many methods fail to solve the problem
of transition from a weak to a
strong coupling limit. The advantage of the ORM
is in the consistent treatment of the problem, while
different approaches run into a problem of
summing of the perturbation series.
In addition, the ground state and the orbital and radial excitations
are treated in a similar way within the ORM.
We note that our results are in the remarkable agreement
with the ones, obtained for example in \cite{Pop}
at different limits.
The ORM  has been applied successfully to
calculate the energy spectrum  of Coulomb and
Yukawa potentials for different sets of their strengths.
The results show  good agreement with the quantitative
solutions of the problem \cite{Ad}.
In fact, in both cases our solutions have been obtained
avoiding the question of the Dyson phenomenon \cite{Dy}
which causes the main difficulties in the perturbation approach.
We hope that the results obtained for the
considered problems can be useful
for an analysis of different questions arising,
in particular, in atomic physics.

In conclusion, the ORM allows in many cases to obtain
analytical results, which are available only numerically or
partially analytically within the perturbation approach or
the WKB approximation.
It is clear, that any progress in the development of analytical methods
can provide a deeper insight into solution and understanding
of the quantum mechanical problem, making further contribution
to quantum mechanics.

\appendix
\def\theequation{\thesection.\arabic{equation}}
\section{Calculation of some products of the annihilation and
creation operators}
\setcounter{equation}{0}

The canonical variables  $q_j$ and $p_j$ are coordinates and
momentum, respectively, in the d-space $R^d$ and are expressed
through $a^+_j$ and $a_j$
\begin{eqnarray}
\label{m1}
q_j=\frac{1}{\sqrt{2\omega}}\left(a_j+a^+_j\right)~,~~~~~~~~~~~~~
p_j=\sqrt{\frac{\omega}{2}}\cdot\frac{a_j-a^+_j}{i}~.
\end{eqnarray}
The operators  $a^+_j$ and  $a_j$ fulfill the standard commutation relation
\begin{eqnarray}
\label{m2}
\left[a_i,a^+_j\right]=\delta_{i,j}~,~~~~~j=1,2,...d~.
\end{eqnarray}
Hereafter $d\equiv D$.
From (\ref{m1}) and (\ref{m2}) we can obtain for $q^2$, $p^2$
\begin{eqnarray}
\label{m3}
q^2=\frac{d}{2\omega}+:q^2:~,~~~~
p^2=\frac{d\omega}{2}+:p^2:~,
\end{eqnarray}
where the symbol $:*:$ means the normal ordering and
$\omega$ is the oscillator frequency.

For $a^+_j$ and $a_j$ we have
\begin{eqnarray}
\label{a1}
e^{i\vec{k}\vec{a}}e^{i\vec{p}\vec{a^+}}=e^{i\vec{p}\vec{a^+}}
e^{i\vec{k}\vec{a}}\cdot e^{-(kp)}~,
\end{eqnarray}
where $\vec{k}$ and  $\vec{p}$ are vectors in the $d$-space.

Let consider the expression
\begin{eqnarray}
\label{a2}
Y_j(\vec{k})=
e^{i\vec{k}\vec{a}}a^+_je^{-i\vec{k}\vec{a}}~.
\end{eqnarray}
At $\vec{k}=0$ from  Eq.(\ref{a2}) we have
\begin{eqnarray}
\label{a3}
Y_j(0)=a^+_j~.
\end{eqnarray}
Taking into account (\ref{m2}), we obtain from (\ref{a2})
\begin{eqnarray}
\label{a4}
\frac{dY_j(\vec{k})}{dk_l}=i\delta_{jl}
\end{eqnarray}
Integrating over $k_l$ and taking into account (\ref{a3}), we have
\begin{eqnarray}
\label{a5}
Y_j(\vec{k})=
e^{i\vec{k}\vec{a}}a^+_je^{-i\vec{k}\vec{a}}=a^+_j+ik_j~.
\end{eqnarray}
Similar expressions can be obtained for
\begin{eqnarray}
\label{a6}
&&e^{-i\vec{p}\vec{a^+}}a_je^{i\vec{p}\vec{a^+}}=a_j+ip_j~,\\
&&e^{\alpha\vec{a^+}\vec{a}}a_je^{-\alpha\vec{a^+}\vec{a}}=a_je^{-\alpha}
\nonumber\\
&&e^{\alpha\vec{a^+}\vec{a}}a^+_je^{-\alpha\vec{a^+}\vec{a}}=a^+_je^{\alpha}~.
\nonumber
\end{eqnarray}

According to the ORM, we transform the Hamiltonian rewriting
all variables into the $d$--space in a normal order.
As a result, the potential can be expressed
as a polynomial of different powers of $q$.
Some typical cases of potentials are the following:

a) a power-law potential
\begin{eqnarray}
\label{m4}
&&q^{2n}=(-1)^n\frac{d^n}{dx^n}\cdot e^{-xq^2}~_{|x=0}\\
&&=(-1)^n\frac{d^n}{dx^n}\cdot\int\left(\frac{d\eta}{\sqrt{\pi}}\right)^d
e^{-\eta^2(1+x/\omega)}:e^{-2i\sqrt{x}(q\eta)}:_{|x=0} \nonumber\\
&&=\frac{1}{\omega^n}\cdot\frac{\Gamma(\frac{d}{2}+n)}{\Gamma(\frac{d}{2})}
+:q^2:\frac{n}{\omega^{n-1}}
\cdot\frac{\Gamma(\frac{d}{2}+n)}{\Gamma(\frac{d}{2}+1)}\nonumber\\
&&+\frac{(-1)^n}{\omega^n}\frac{d^n}{dx^n}\cdot\int\left(\frac{d\eta}{\sqrt{\pi}}\right)^d
e^{-\eta^2(1+x)}:e_2^{-2i\sqrt{x\omega}(q\eta)}:_{|x=0}
\nonumber
\end{eqnarray}
where $n=1,2,...$  are integer and positive numbers.

b) an inverse power-law potential
\begin{eqnarray}
\label{m5}
&&q^{2\tau}=\int\limits_{0}^{\infty}\frac{dx}{\Gamma(-\tau)}
x^{-1-\tau}\cdot e^{-xq^2}\\
&&=\int\limits_{0}^{\infty}\frac{dx}{\Gamma(-\tau)}x^{-1-\tau}
\cdot\int\left(\frac{d\eta}{\sqrt{\pi}}\right)^d
e^{-\eta^2(1+x/\omega)}:e^{-2i\sqrt{x}(q\eta)}:\nonumber\\
&&=\frac{1}{\omega^\tau}\cdot\frac{\Gamma(\frac{d}{2}+\tau)}{\Gamma(\frac{d}{2})}
+:q^2:\frac{\tau}{\omega^{\tau-1}}
\cdot\frac{\Gamma(\frac{d}{2}+\tau)}{\Gamma(\frac{d}{2}+1)}\nonumber\\
&&+\frac{1}{\omega^\tau}\int\limits_{0}^{\infty}\frac{dx}{\Gamma(-\tau)}x^{-1-\tau}
\cdot\int\left(\frac{d\eta}{\sqrt{\pi}}\right)^d
e^{-\eta^2(1+x)}:e_2^{-2i\sqrt{x\omega}(q\eta)}:
\nonumber
\end{eqnarray}
Here $\tau\neq n$; we use the notation $e_2^x=e^x-1-x-x^2/2$.

c) In the general case we use the Fourier transform
\begin{eqnarray}
\label{m6}
W(q^2)&=&\int{\Big(}{dk\over2\pi}{\Big)}^d\tilde{W}_d(k^2)e^{ikq}\\
&=&
\int{\Big(}{dk\over2\pi}{\Big)}^d\tilde{W}_d(k^2)
\exp{\Big (}ik{a+a^+\over\sqrt{2\omega}}{\Big )}
\nonumber\\
&=&\int{\Big(}{dk\over2\pi}{\Big)}^d\tilde{W}_d(k^2)
\exp{\Big (}-{k^2\over{4\omega}}{\Big )}
\exp{\Big (}ik{a^+\over\sqrt{2\omega}}{\Big )}
\exp{\Big (} ik{a\over\sqrt{2\omega}}{\Big )}\nonumber\\
&=&\int{\Big(}{dk\over2\pi}{\Big)}^d\tilde{W}_d(k^2)
\exp{\Big (}-{k^2\over{4\omega}}{\Big )}:\exp(ikq):,
\nonumber
\end{eqnarray}
where $(kq)=\sum k_jq_j$ and
\begin{eqnarray}
{\widetilde{W}}_d(k^2)=\int(d\rho)^dW(\rho^2){\rm e}^{i(k\rho)}~.
\nonumber
\end{eqnarray}

The normal order
form of an arbitrary potential can be defined by
Eqs.(\ref{m4})-(\ref{m6}). In particular, we have

for $n=1,~2,~3$
\begin{eqnarray}
\label{m7}
&&q^2=\frac{d}{2\omega}+:q^2:~,\\
&&q^4=\frac{d(d+2)}{4\omega^2}+\frac{d+2}{\omega}:q^2:+q^4:~,\nonumber\\
&&q^6=\frac{d(d+2)(d+4)}{8\omega^3}+
\frac{3(d+2)(d+4)}{4\omega^2}:q^2:
+\frac{3(d+4)}{2\omega}:q^4:+:q^6:~.\nonumber
\end{eqnarray}

For a calculation of the energy and different matrix elements
we need to know the result of action of
a combination $\left(a^+a^+\right)^n$ or $(aa)^n$.
The following representation is useful
\begin{eqnarray}
\label{m8}
&&\left(a^+a^+\right)^n=(-1)^n\frac{d^n}{d\beta^n}\cdot
e^{-\beta(a^+a^+)}~{\bigg|}_{\beta=0}\\
&&=(-1)^n\frac{d^n}{d\beta^n}\cdot
\int\left(\frac{d\zeta}{\sqrt{\pi}}\right)^d
e^{-\zeta^2-2i\sqrt{\beta}(a^+\zeta)}{\bigg|}_{\beta=0}~.
\nonumber
\end{eqnarray}

According to Eqs.(\ref{m4})-(\ref{m6}), the interaction Hamiltonian
contains the term proportional to $\exp_2\{-i(\vec{k}\vec{a})\}$.
Let us introduce the following operators:
\begin{eqnarray}
\label{e7}
e_2^{-i(\vec{k}\vec{a})}={\sl P}_\nu\cdot
e^{-i\nu(\vec{k}\vec{a})}~,
\end{eqnarray}
where ${\sl P}_\nu$ is the operator defined
according to the following rules:
\begin{eqnarray}
\label{e8}
&&{\sl P}_\nu\cdot Const=0\\
&&{\sl P}_\nu\cdot \nu^n=0~,~~~{\rm ¯at~~~ n\leq2}\nonumber\\
&&{\sl P}_\nu\cdot \nu^n=1~,~~~{\rm ¯at~~~ n>2}\nonumber
\end{eqnarray}

\section{Normalization}
\setcounter{equation}{0}

In the ORM, the wave function is defined
\begin{eqnarray}
\label{n1}
\ket{n_r} = C_{n_r}\left(a^+_ja^+_j\right)^{n_r}\ket{0}~,
~~~~~~~j=1,...,d,
\end{eqnarray}
where $C_{n_r}$ is the normalization constant. Hereafter we
use the notation $n\equiv n_r$.
The constant can be determined from the condition
\begin{eqnarray}
\label{n2}
1=<n\ket{n}=C^2_n\bra{0}\left(a_ia_i\right)^n
\left(a^+_ja^+_j\right)^n\ket{0}~.
\end{eqnarray}
Using the definition (\ref{n1}),  the representation (\ref{m4}),
and the normalization condition $<0\ket{0}=1$ from (\ref{n2}) we obtain
\begin{eqnarray}
\label{n3}
&&1=C^2_n\frac{\partial^{2n}}{\partial\alpha^n\partial\beta^n}\cdot
\int\left(\frac{d\eta}{\sqrt{\pi}}\right)^d
\int\left(\frac{d\xi}{\sqrt{\pi}}\right)^d
e^{-\eta^2-\xi^2}\cdot \\
&&\times\bra{0}e^{-2i\sqrt{\alpha}(a\xi)} \cdot
e^{-2i\sqrt{\beta}(a^+\eta)}\ket{0}{\bigg|}_{\alpha,\beta=0}
\nonumber\\
&&=C^2_n\frac{\partial^{2n}}{\partial\alpha^n\partial\beta^n}\cdot
\int\left(\frac{d\eta}{\sqrt{\pi}}\right)^d
\int\left(\frac{d\xi}{\sqrt{\pi}}\right)^d
e^{-\eta^2-\xi^2-4\sqrt{\alpha\beta}(\xi\eta)}
{\bigg|}_{\alpha,\beta=0}
\nonumber\\
&&=C^2_n\frac{\partial^{2n}}{\partial\alpha^n\partial\beta^n}\cdot
\frac{1}{(1-4\alpha\beta)^{d/2}}
{\bigg|}_{\alpha,\beta=0}~.
\nonumber
\end{eqnarray}
Finally, for $C_n$ we can write
\begin{eqnarray}
\label{n4}
C_n=\left(\frac{\Gamma(d/2)}{4^nn!\Gamma(d/2+n)}
\right)^{1/2}~.
\end{eqnarray}

Let consider the result of the action of some operators
on the wave function (\ref{n1}).
By definition
($\bra{0}a^+=0,$ $a\ket{0}=0,$ where  $\bra{0}0\rangle=1$),
it follows that
\begin{eqnarray}
\label{n5}
&&e^{-B(a^+a)}\ket{n}=
C_ne^{-B(a^+a)}\left(a^+a^+\right)^n\ket{0}\\
&&=C_n(-1)^n\frac{\partial^{n}}{\partial\beta^n}\cdot
\int\left(\frac{d\eta}{\sqrt{\pi}}\right)^d
e^{-\eta^2}e^{-B(a^+a)}
e^{-2i\sqrt{\beta}(a^+\eta)}\ket{0}
{\bigg|}_{\beta=0}
\nonumber\\
&&=C_n(-1)^n\frac{\partial^{n}}{\partial\beta^n}\cdot
e^{-\beta e^{-2B}}\ket{0}
{\bigg|}_{\beta=0}
=e^{-2nB}\ket{n}\nonumber
\end{eqnarray}

\section{Calculation of corrections to the energy spectrum}
\setcounter{equation}{0}

In the ORM the Hamiltonian has the form
\begin{eqnarray}
\label{f1}
H=H_0+H_I+\varepsilon_0~.
\end{eqnarray}
The contribution of the interaction Hamiltonian $H_I$ is
considered in the perturbation approach.
The energy and  wave function are defined
\begin{eqnarray}
\label{f2}
&&E_n=E^{(0)}_0+E^{(0)}_n+E^{(1)}_n+E^{(2)}_n+\cdots \\
&&\Psi_n=\Psi^{(0)}_n+\Psi^{(1)}_n+\Psi^{(2)}_n+
\cdots~, \nonumber
\end{eqnarray}
where $E^{(0)}_0=\varepsilon_0$ is the ground state energy.
Within the perturbation approach the SE has the form
\begin{eqnarray}
\label{f3}
H_0\Psi^{(k)}_n+H_I\Psi^{(k-1)}_n=
\sum_{k_1+k_2=k}E^{(k_1)}_n\Psi^{(k_2)}_n~,
\end{eqnarray}
where  n is a radial quantum number and k is an order of the
perturbation approach.
In the zeroth approximation, we have from Eq.(\ref{f3})
\begin{eqnarray}
\label{f4}
H_0\Psi^{(0)}_n=E^{(0)}_n\Psi^{(0)}_n~.
\end{eqnarray}
According to the ORM, $H_0=\omega(a^+a)$ and
$\Psi^{(0)}_n=|n>$. From Eq.(\ref{f4}) it follows that
\begin{eqnarray}
\label{f5}
&&E^{(0)}_n=\left(\Psi^{(0)}_nH_0\Psi^{(0)}_n\right)=
\bra{n}\omega(a^+a)\ket{n}~,\\
&&H_0=\omega(a^+a)~,~~~~~~\Psi^{(0)}_n=\ket{n}~.\nonumber
\end{eqnarray}
For k=1 we obtain from Eq.(\ref{f3})
\begin{eqnarray}
\label{f6}
H_0\Psi^{(1)}_n+H_I\Psi^{(0)}_n=
E^{(0)}_n\Psi^{(1)}_n+E^{(1)}_n\Psi^{(0)}_n~.
\end{eqnarray}
Taking into account Eq.(\ref{f5}), we obtain from Eq.(\ref{f6})
\begin{eqnarray}
\label{f7}
&&E^{(1)}_n = \left(\Psi^{(0)}_nH_I\Psi^{(0)}_n\right)~,\\
&&\Psi^{(1)}_n=-\frac{1}{H_0-2n\omega}
\left[H_I-E^{(1)}_n\right]\Psi^{(0)}_n~. \nonumber
\end{eqnarray}
Following the same procedure, from Eq.(\ref{f3}) we obtain
\begin{eqnarray}
\label{f8}
&&E^{(2)}_n =-\left(\Psi^{(0)}_n
\left[H_I-E^{(1)}_n\right]\frac{1}{H_0-2n\omega}
\left[H_I-E^{(1)}_n\right]\Psi^{(0)}_n\right)~.
\end{eqnarray}
\begin{itemize}

\item {\bf Calculation of $E^{(0)}_n$.}

Taking into account Eq.(\ref{n1}) and  Eq.(\ref{m8}),
we obtain from Eq.(\ref{f5})
\begin{eqnarray}
\label{en1}
&&E^{(0)}_n=\omega C^2_n\bra{0}(a_ia_i)^n(a^+_ja_j)(a^+_la^+_l)^n\ket{0}
=\omega C^2_n
\frac{\partial^{2n}}{\partial\alpha^n\partial\beta^n}\\
&&\int\left(\frac{d\eta}{\sqrt{\pi}}\right)^d
\int\left(\frac{d\xi}{\sqrt{\pi}}\right)^d
e^{-\eta^2-\xi^2}
\bra{0}e^{-2i\sqrt{\alpha}(a\xi)} \left(a^+_ja_j\right)
e^{-2i\sqrt{\beta}(a^+\eta)}\ket{0}{\bigg|}_{\alpha,\beta=0}
\nonumber
\end{eqnarray}
Using Eqs.(\ref{a1}), (\ref{a5}) and (\ref{a6}), after
some manipulations we have
\begin{eqnarray}
\label{en2}
&&E^{(0)}_n
=\omega C^2_n \frac{\partial^{2n}}{\partial\alpha^n\partial\beta^n}
\cdot \frac{\partial}{\partial\tau}
\int\left(\frac{d\eta}{\sqrt{\pi}}\right)^d
\int\left(\frac{d\xi}{\sqrt{\pi}}\right)^d
e^{-\eta^2-\xi^2-4\tau\sqrt{\alpha\beta}(\xi\eta)}
{\bigg|}_{\alpha,\beta=0;~\tau=1}\nonumber\\
&&
=\omega C^2_n \frac{\partial^{2n}}{\partial\alpha^n\partial\beta^n}
\cdot \frac{\partial}{\partial\tau}
\frac{1}{\left(1-4\tau^2\alpha\beta\right)^{d/2}}
{\bigg|}_{\alpha,\beta=0;~\tau=1}~.
\end{eqnarray}
Finally, we obtain from Eq.(\ref{en2})
\begin{eqnarray}
\label{en3}
&&E^{(0)}_n=\bra{n}H_0\ket{n}=2n\omega~.
\end{eqnarray}

In order to demonstrate the principles of a calculation of the first
order, let us consider the interaction Hamiltonian in the
form
\begin{eqnarray}
\label{in1}
H_I=g\frac{(-1)^\tau}{\omega^\tau}
\frac{\partial^\tau}{\partial x^\tau}
\cdot \int\left(\frac{d\eta}{\sqrt{\pi}}\right)^d
e^{-\eta^2(1+x)}:e_2^{-2i\sqrt{x\omega}(q\eta)}:
{\bigg|}_{x=0}~,
\end{eqnarray}
where  $g$ is the interaction constant and
$\tau $ is some constant.

\item {\bf Calculation of $E^{(1)}_n$.}

Taking into account Eq.(\ref{in1}),  we obtain from (\ref{f7})
\begin{eqnarray}
\label{in2}
E^{(1)}_n=g\frac{(-1)^\tau}{\omega^\tau}
\frac{\partial^\tau}{\partial x^\tau}
\cdot \int\left(\frac{d\eta}{\sqrt{\pi}}\right)^d
e^{-\eta^2(1+x)}\bra{n}:e_2^{-2i\sqrt{x\omega}(q\eta)}:
\ket{n}{\bigg|}_{x=0}~
\end{eqnarray}
For calculations of the first order contributions,
we very often treat  the following term
\begin{eqnarray}
\label{in3}
T_n(x)=\int\left(\frac{d\eta}{\sqrt{\pi}}\right)^d
e^{-\eta^2(1+x)}\bra{n}:e_2^{-2i\sqrt{x\omega}(q\eta)}:
\ket{n}.
\end{eqnarray}
Let us consider in more detail the procedure of calculating
this term. Taking into account Eqs.(\ref{e7}), (\ref{n1}), (\ref{m8}),
we obtain  for Eq.(\ref{in3})
\begin{eqnarray}
\label{in4}
&&T_n(x)={\sl P}_\nu\int\left(\frac{d\eta}{\sqrt{\pi}}\right)^d
e^{-\eta^2(1+x)}\bra{n}:e^{-2i\nu\sqrt{x\omega}(q\eta)}:
\ket{n}\\
&&={\sl P}_\nu C_n^2
\frac{\partial^{2n}}{\partial\alpha^n\partial\beta^n}
\int\left(\frac{d\eta}{\sqrt{\pi}}\right)^d
\int\left(\frac{d\xi_1}{\sqrt{\pi}}\right)^d
\int\left(\frac{d\xi_2}{\sqrt{\pi}}\right)^d
e^{-\eta^2(1+x)-\xi^2_1+\xi^2_2}\nonumber\\
&&\bra{0}e^{-2i\sqrt{\alpha}(a\xi_1)}
e^{-i\nu\sqrt{2x}(a^+\eta)} e^{-i\nu\sqrt{2x}(a\eta)}
e^{-2i\sqrt{\beta}(a^+\xi_2)}\ket{0}{\Bigg|}_{\alpha,\beta=0}
\nonumber
\end{eqnarray}
Finally, we have
\begin{eqnarray}
\label{in5}
T_n(x)&=&\sum_{k=2}^{2n}\sum_{s=0}^{n}(-1)^k
\frac{x^k}{(1+x)^{k+d/2}}\frac{\Gamma(1+n)}{\Gamma(n+d/2)}
\frac{2^{2s-k}}{\Gamma(n-s+1)} \\
&&\frac{\Gamma(k+n-s+d/2)}
{\Gamma^2(k-s+1)\Gamma(2s-k+1)}~.
\nonumber
\end{eqnarray}
According to Eq.(\ref{in2}), the first order correction
to the energy of a state is
\begin{eqnarray}
\label{in6}
E^{(1)}_n=g\frac{(-1)^\tau}{\omega^\tau}
\frac{\partial^\tau}{\partial x^\tau}
T_n(x)
{\bigg|}_{x=0}~.
\end{eqnarray}

Also, in different cases it is useful to know the result of
calculations of the following quantity
\begin{eqnarray}
\label{in7}
Q_n(\tau)=\int\limits_{0}^{\infty}\frac{dx}{\Gamma(-\tau)}
x^{-1-\tau}\cdot T_n(x)~.
\end{eqnarray}
Taking into account Eq.(\ref{in5}), after integration
over $x$ and some manipulation we obtain for Eq.(\ref{in7})
\begin{eqnarray}
\label{in8}
Q_n(\tau)&=&\frac{\Gamma(d/2-\tau)}{\Gamma(\tau)}
\frac{\Gamma(1+n)}{\Gamma(n+d/2)}
\sum_{k=2}^{2n}
\frac{\Gamma(\tau+k)}{\Gamma(k+d/2)} \\
&&\sum_{s=0}^{n}(-1)^k
\frac{2^{2s-k}}{\Gamma(n-s+1)}\frac{\Gamma(k+n-s+d/2)}
{\Gamma^2(k-s+1)\Gamma(2s-k+1)}~.
\nonumber
\end{eqnarray}
Eqs.(\ref{in5})  and (\ref{in7})
allow one to calculate any potential matrix elements.
In particular, using Eq.(\ref{in5}),
we have very simple expressions for the power-law potential
\begin{eqnarray}
\label{in9}
&&\bra{n_r}:q^2:\ket{n_r}=\frac{2n_r}{\omega}~\\
&&\bra{n_r}:q^4:\ket{n_r}=\frac{n_r}{\omega^2}\left[d+6n_r-4\right]~,
\nonumber\\
&&\bra{n_r}:q^6:\ket{n_r}=\frac{2n_r(n_r-1)}{\omega^3}
\left[3d+10n_r-8\right]~.
\nonumber
\end{eqnarray}
\end{itemize}


\newpage
\begin{center}
\begin{tabular}{|l c llll|}
\hline
 & &~\cite{Pop1}&\cite{Cor}
&\cite{Pop} &${\rm ORM}$\\
\hline
$g=0.976562$ & $\ell=0$ & 0.556764
                   & 0.556767
                   & 0.557
                   & 0.5580 \\
           & $\ell=1$ &&&& 1.9375\\
$g=4.0$      & $\ell=0$ &2.79573
                   &2.79575
                   &2.796
                   & 2.7972\\
           & $\ell=1$ &&&& 5.65135\\
$g=62.5$     & $\ell=0$ &24.79
                   &24.856
                   &24.86
                   &24.8569\\
           & $\ell=1$ &&&&39.4096\\
$g=100$      & $\ell=0$ &34.75
           &        &34.9
                    &34.9049\\
           & $\ell=1$ &&&&54.4364\\
$g=500$      & $\ell=0$ &107.0
                   &108.366
                   &108.4
                   &108.3854\\
           & $\ell=1$ &&&&162.898\\
$g=1000$     & $\ell=0$ &174.8&
           &        &174.8698\\
           & $\ell=1$ &&&&260.295\\
$g\to\infty$ & $\ell=0$ &$C=1.764$
                   &$1.85576$
                   &$1.856$
                   &$1.8559$\\
           & $\ell=1$ &&&&$0.668$\\
           & $\ell=3$ &&&&$0.009$\\
\hline
\end{tabular}
\end{center}
\begin{center}
{\bf Table 1.} The energy of the ground state and orbital excitations
\end{center}

\newpage
\begin{center}
\begin{tabular}{|l|ll ll ll|}
\hline
 & \multicolumn{2}{|c}{$\ell=0$}
 & \multicolumn{2}{|c|}{$\ell=1$}
 & \multicolumn{2}{c|}{$\ell=2$}  \\
\cline{2-3} \cline{4-5} \cline{6-7}

$c$&$\rho$& $-E$ &$\rho$& $-E$ &$\rho$& $-E$ \\
\hline
0.001  & 1.0  &  0.251  & 1.0 &0.06350&1.0&0.02877 \\
       &      &  (0.251)&    &(0.06350)& &(0.02877) \\
0.005  & 1.0  &  0.25497  & 1.0 &0.06740&1.0&0.032525 \\
       &      &  (0.25496)&    &(0.06738)& &(0.032520) \\
0.01   & 1.0  &  0.25985  & 1.0 &0.07204&0.99&0.036813 \\
       &      &  (0.25895)&    &(0.07202)& &(0.036810) \\
0.05   & 1.0  &  0.29650  & 0.99&0.10219&0.91&0.060521 \\
       &      & (0.29650)&    &(0.10242& &(0.060520) \\
0.1    & 0.99  & 0.33694  & 0.91 &0.12931&0.88&0.077046 \\
       &      & (0.33694)&    &(0.12931)& &(0.077050) \\
0.2    & 0.97  & 0.404244  & 0.88 &0.164292&0.88&0.093731 \\
       &      &  (0.40424)&    &(0.16429)& &(0.093730) \\
0.5    & 0.91  &  0.54246  & 0.87 &0.211998&0.93&0.107695 \\
       &      &  (0.54243)&    &(0.21203)& &(0.10770) \\
1.0    & 0.89  & 0.67482  & 0.89 &0.236543&.98&0.110671 \\
       &       & (0.67482)&    &(0.23665)& &(0.11067) \\
2.0    & 0.89  &  0.80554 & 0.97 &0.24709&1.00&0.111084 \\
       &      &  (0.80566)&    &(0.24711)& &(0.11108) \\
10.0   & 0.90  &  0.97223  & 1.00 &0.249996&1.00&0.111111 \\
       &      &  (0.97424)&    &(0.24998)& &(0.11111) \\
\hline
\end{tabular}
\end{center}
\begin{center}
{\bf Table 2.} Energy of ESCP for $B=1$ as a function of the
 parameter $c$. Result of calculations \cite{Ad} are presented in brackets.
\end{center}

\newpage
\begin{center}
\begin{tabular}{|l| ll ll ll|}
\hline
 & \multicolumn{2}{|c}{$\ell=0$}
 & \multicolumn{2}{|c|}{$\ell=1$}
 & \multicolumn{2}{c|}{$\ell=2$}  \\
\cline{2-3} \cline{4-5} \cline{6-7}

$c$&$\rho$& $-E$ &$\rho$& $-E$ &$\rho$& $-E$ \\
\hline
0.001  & 1.0  &  0.00168  & 1.0 &0.00160&1.0&0.001538 \\
       &      &  (0.00178)&    &(0.00168)& &(0.00160) \\

0.005  & 0.96  &  0.00807  & 0.87 &0.00735&0.88&0.006746 \\
       &      &  (0.00814)&    &(0.00739)& &(0.00678) \\

0.01   & 0.88  &  0.015264  & 0.72 &0.013575&0.755&0.012125 \\
       &      &  (0.01540)&    &(0.01360)& &(0.01215) \\

0.05   & 0.69 &  0.063234  & 0.655&0.050177&0.695&0.040366 \\
       &      & (0.06325&         &(0.05019)& &(0.04039) \\

0.1    & 0.63  & 0.11136  & 0.67 &0.081862&0.71&0.060999 \\
       &      & (0.11130)&       &(0.08188)& &(0.06100) \\

0.2    & 0.64  & 0.1885  & 0.705 &0.12482&0.77&0.083568 \\
       &      & (0.18847)&    &(0.12486)& &(0.08360) \\

0.5    & 0.69  &  0.34885  & 0.755 &0.18899&0.88&0.104978 \\
       &      &  (0.34891)&    &(0.18915)& &(0.10502) \\

1.0    & 0.72  & 0.51243  & 0.86 &0.226511&.965&0.110258 \\
       &       & (0.51246)&    &(0.22668)& &(0.11027) \\

2.0    & 0.7863  &  0.6877 & 0.94 &0.244485&0.995&0.111057 \\
       &      &  (0.68856)&    &(0.24458)& &(0.11106) \\

10.0   & 0.93  &  0.95111  & 1.00 &0.24999&1.00&0.11111 \\
       &      &  (0.95197)&    &(0.24997)& &(0.11111) \\
\hline
\end{tabular}
\end{center}
\begin{center}
{\bf Table 3.} Similar to Table.2 for $B=2$.
\end{center}

\begin{center}
\begin{tabular}{|l|ll ll ll|}
\hline
 & \multicolumn{2}{|c}{$\ell=0$}
 & \multicolumn{2}{|c|}{$\ell=1$}
 & \multicolumn{2}{c|}{$\ell=2$}  \\
\cline{2-3} \cline{4-5} \cline{6-7}
$c$&$\rho$& $-E$ &$\rho$& $-E$ &$\rho$& $-E$ \\
\hline
0.005  & 0.815  &  0.0033  & 0.495 &0.0034&0.455&0.00345 \\
       &      &  (0.00359) &    &(0.00355)& &(0.00351) \\
0.01   & 0.50  &  0.0062 & 0.465 &0.0065&0.655&0.0066 \\
       &      &  (0.00700)&    &(0.00694)& &(0.00681) \\
0.05   & 0.50  &  0.0311  & 0.465&0.0305&0.53&0.02835 \\
       &      & (0.03235)&    &(0.03090)& &(0.02840) \\
0.1    & 0.51  & 0.0605  & 0.415 &0.0555&0.535&0.04780 \\
       &      & (0.06088)&    &(0.05570)& &(0.04782) \\
0.2    & 0.52  & 0.110  & 0.48 &0.0943&0.625&0.07261 \\
       &      &  (0.11160)&    &(0.09451)& &(0.07260) \\
0.5    & 0.52  &  0.233  & 0.64 &0.16370&0.81&0.10098 \\
       &      &  (0.23359)&    &(0.16380)& &(0.10098) \\
1.0    & 0.53  & 0.3790  & 0.77 &0.21265&.94&0.10951 \\
       &       & (0.37984)&    &(0.21279)& &(0.10954) \\
2.0    & 0.65  &  0.56403 & 0.90 &0.2402&0.995&0.11104 \\
       &      &  (0.56407)&    &(0.24029)& &(0.11101) \\
10.0   & 0.94  &  0.91317  & 1.00 &0.250&0.667&0.11111 \\
       &      &  (0.91540)&    &(0.24993)& &(0.11111) \\
\hline
\end{tabular}
\end{center}

\begin{center}
{\bf Table 4.} Similar to Table.2 for $B=4$.
\end{center}
\end{document}